\documentclass[12pt,preprint]{aastex}
\usepackage{natbib}
\usepackage{epsf}
\usepackage[dvips]{epsfig}
\usepackage{times}

\shorttitle{Star Cluster Survival}
\shortauthors{Fellhauer \& Kroupa}

\begin{document}

\title{Star Cluster Survival in Star Cluster Complexes under Extreme
  Residual Gas Expulsion} 

\author{M. Fellhauer and P. Kroupa}
\affil{Sternwarte University of Bonn, 53121 Bonn, Germany}
\email{mike,pavel@astro.uni-bonn.de}

\begin{abstract}
  After the stars of a new, embedded star cluster have formed they
  blow the remaining gas out of the cluster.  Especially winds of
  massive stars and definitely the on-set of the first supernovae can
  remove the residual gas from a cluster.  This leads to a very violent
  mass-loss and leaves the cluster out of dynamical equilibrium.
  Standard models predict that within the cluster volume the star
  formation efficiency (SFE) has to be about $33$~per cent for sudden
  (within one crossing-time of the cluster) gas expulsion to retain
  some of the stars in a bound cluster.  If the efficiency is lower
  the stars of the cluster disperse mostly.  

  Recent observations reveal that in strong star bursts star clusters
  do not form in isolation but in complexes containing dozens and up
  to several hundred star clusters, i.e.\ in super-clusters.  By
  carrying out numerical experiments for such objects placed at
  distances $\geq 10$~kpc from the centre of the galaxy we demonstrate
  that under these conditions (i.e.\ the deeper potential of the star
  cluster complex and the merging process of the star clusters within
  these super-clusters) the SFEs can be as low as $20$~per cent and
  still leave a gravitationally bound stellar population.  Such an
  object resembles the outer Milky Way globular clusters and the faint
  fuzzy star clusters recently discovered in NGC~1023.
\end{abstract}


\maketitle

\section{Introduction}
\label{sec:intro}

Star clusters form out of collapsing cloud clumps in molecular clouds
\citep{til04}.  These collapses are triggered by turbulent
fragmentation of clouds and their clumps \citep{mac04}.  The clumps
are observed to be aligned in filamentary structures that can be
reproduced by supersonic turbulent simulations and they contain many
cores \citep{bur00,kle01}.  Each core forms a single star or a
binary.  The mass function found for these cores is essentially the
same as the initial mass function of the stars \citep{joh00}.  While
the star formation efficiency (SFE; i.e.\ the fraction of gas which
ends up in the star(s)) in these cores is high, the overall
SFE measured over the whole molecular cloud is very low, of the order
of a few per cent \citep{cla04}, and $\leq 40$~per cent in cluster
forming clumps \citep[i.e.\ embedded star clusters;][]{lad03}. 

The remaining gas does not stay in the newborn star cluster but is
driven outwards by stellar feedback.  In embedded clusters containing
more than a few hundred stars the feedback consists of photo-ionising
radiation, the winds of high-mass stars and finally the on-set
of the first supernova explosions \citep{goo97}.  For such clusters
the feedback energy can easily be sufficient to unbind the gas leading
to a gas expulsion phase which is rather short, comparable to the
crossing time of the star cluster.  Pictures of young massive star
clusters, for example in the central region of the Antennae
\citep[NGC~4038/4039;][]{whi99}, reveal that they are already
surrounded by H-$\alpha$ bubbles stemming from the gas which was blown
out of the star cluster.  It has been shown that these star clusters
can be as young as $5$--$6$~Myr.  The outflow velocities of the gas
has been measured to be $25$--$30$~kms$^{-1}$ \citep{whi99}, which
corresponds to gas-evacuation times of $0.2$~Myr for cluster radii
of $4$~pc.  This is comparable to the crossing time of a
$10^{5}$~M$_{\odot}$ cluster.

As a result of this strong and rapid mass-loss the star cluster is
left out of virial equilibrium.  The velocities of the stars are too
high for the reduced mass of the star cluster.  So even  more mass is
lost when stars escape from the star cluster.  This may finally lead
to the complete dissolution of the cluster.  But if the star formation
efficiency is as high as $33$~per cent or above, a small bound core
remains \citep{goo97, boi03a, boi03b} and today we therefore
understand the formation of low-mass, Pleiades-type clusters
\citep{kro01}.  \citet{gey01} argue that the SFE has to be larger than
$50$~per cent to get a bound core but if the stars have almost no
initial velocity dispersion then $10$~per cent could be enough.  This
means that the measurement of the initial velocity distribution in
newborn embedded star clusters will be crucial to find out if and how
star clusters survive.   

As seen in the beautiful HST images of the central region of the
Antennae galaxies \citep{whi99}, star clusters in strong star bursts
do not form in isolation but rather in star cluster complexes.  These
are confined regions of several hundred parsecs containing dozens and
up to hundreds of young massive star clusters \citep{kro98, zha99}.

Following the arguments of \citet{kro98} these star cluster
complexes are bound entities and should be close to virial
equilibrium.  Even though they are quite young (in the Antennae $\leq
10$~Myr) they should be already more dispersed and not centrally
concentrated anymore if they are unbound, or almost all clusters
should be merged in the centre if they are sub-virial.  Until now
there is no observational measurement of the velocity dispersion in
the cluster complex available.  Observers have only measured the
velocity dispersions of the star clusters.  However, the difference in
radial velocity of two star clusters in the same cluster complex
($\approx 20$~kms$^{-1}$) is in good agreement with the assumption
that the complex is a bound entity.

In this paper we investigate the birth of massive star clusters in
such extreme environments. Our aim is to understand what r\^{o}le
star cluster complexes, with their collective potential and the
ability for the star clusters to merge with each other on short
timescales \citep{fel02}, have for retaining a bound core. 

In the next section we describe the code we are using and the setup of
our models. Then we present our results followed with a discussion.

\section{Setup}
\label{sec:setup}

We use the particle-mesh code {\sc Superbox} \citep{fel00} which
allows us to keep track of many objects in one simulation.  This code
has a hierarchical grid structure where the high-resolution sub-grids
stay focused on the simulated objects while they move through the
simulation area.  Each grid contains $64^{3}$ grid-cells in this
simulation.  This enables us to resolve the forces between the
star clusters as well as the forces within a star cluster correctly.
We choose the grids of {\sc Superbox} in a way that the innermost
grid-level with the highest resolution ($1.0$~pc per cell) covers each
star cluster, while the medium resolution grids ($5.0$~pc per cell)
have the size of the star cluster complex.  Finally the outermost grid
($500$~pc per cell) covers the whole orbit of the star cluster complex
around the host galaxy.  The time-step of our simulations was chosen
to be $0.1$~Myr which ensures enough time-steps per crossing time of
the single clusters ($24$ per crossing-time) and of the dense star
cluster complex ($59$ per crossing-time).  The CPU time needed was
about $96$~sec per time-step, or about $10$~days for a
$1$~Gyr simulation (on standard desktop PCs). 

The single star clusters are represented by Plummer spheres with a
Plummer radius of $4$~pc and a cut-off radius of $25$~pc.  Each
cluster has a mass of $10^{5}$~M$_{\odot}$ initially, a crossing time
of $2.4$~Myr and is represented with 100,000 particles.  

$20$ of these clusters are placed in a star cluster complex which is
modelled again as a Plummer distribution (i.e. positions and
velocities according to the Plummer distribution function), now with
the star clusters as 'particles'.  This Plummer distribution is given
a Plummer radius of $20$~pc, a cut-off radius of $100$~pc, a crossing
time of $5.9$~Myr and a characteristic velocity dispersion of
$11.25$~kms$^{-1}$.  This is a very dense configuration which will
lead to a fast merging of the star clusters into one massive merger
object, if mass-loss is not taken into account.  It also implies that
the star clusters as well as the star cluster complex is in virial
equilibrium before the gas-expulsion.

\begin{figure}[h!]
  \centering
  \plotone{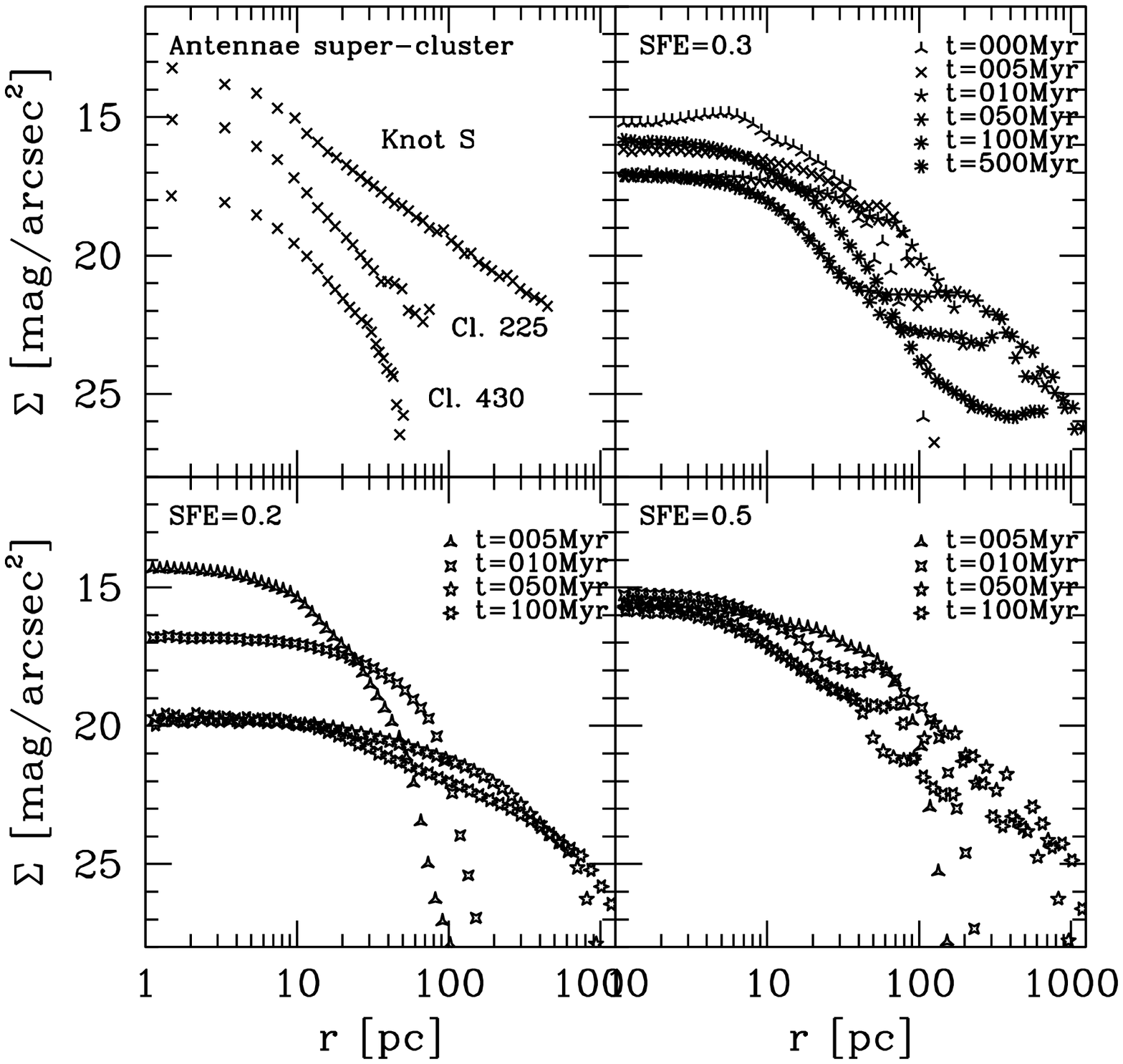}
  \caption{Surface-brightness profiles. Upper left: Star cluster
    complexes in the Antennae. Knot~S and Cluster~225 are young (age
    $<10$~Myr), while Cluster~430 is of intermediate age (a few
    $100$~Myr).  Data are taken from \citet{whi99}. Upper right:
    Model star cluster complex or merger object, respectively, with a
    SFE of $0.3$ (coeval) at different times.  The bumps and
    wiggles in the profile at $t=0$ are due to the individual star
    clusters.  Also visible is the decrease and later increase again
    of the central brightness. This is due to the fact that the
    complex and the merger object first expands and the bound mass
    later contracts again.  Lower left: Model star cluster complex or
    merger object, respectively, with a SFE of $0.2$ (delayed).  Here
    the merger object stays at a low brightness with an extended core.
    Lower right: At a SFE of $0.5$ (coeval) almost no change in the
    central profile is found.  In all cases the merger object fills
    its tidal radius at later times.} 
  \label{fig:compare}
\end{figure}

The initial conditions of our cluster complex are comparable to the
cluster complexes in the central regions of the Antennae galaxies only
in a qualitative way.  However, cluster complexes are not only found
in the central star bursts of interacting galaxies like the Antennae,
but also in tidal tails (e.g.\ Tadpole galaxies, Stephan's Quintet and
also the Antennae) and in more quiescent galaxies like NGC~6946
\citep{lar02}.  They cover a wide range in total mass
($10^{6}$--$10^{8}$~M$_{\odot}$ or even higher) and central
concentrations.  The knots in the Antennae galaxies are much more
massive than our initial conditions and also cover a larger area.
Nevertheless the density distribution there is exponential in the
inner part and has a power-law drop off in the outer part
\citep{whi99}.  Our models, especially after the merging of the first 
few clusters show exactly the same behaviour \citep{fel02b}.  In
Fig.~\ref{fig:compare} we compare the surface-brightness profiles of
our models (measured at different times using mass-to-light ratios
from a single stellar population computed with Starburst99
\citep{lei99}) with the profiles found by \citet{whi99} for three
super-clusters in the Antennae. 

A detailed discussion about merging time-scales, the effect of tides
on the formation of the merging object as well as a discussion of the
properties of the merger objects (without mass-loss) in general is
found in our previous papers \citep{fel02,fel02a,fel02b}.

For our simulations in this project we choose a very dense
configuration for the star cluster complex.  In this case the merging
time-scale becomes comparable to the mass-loss time-scale and the
effect of the cluster complex on retaining stars forming an extended
merger object should be strongest.  The SFE is varied over a wide
range to investigate the influence of this new environment
theoretically.

The mass-loss due to gas-expulsion is modelled by all particles
loosing a fraction of their mass linearly over a crossing-time of the
single star cluster.  We consider two mass-loss models.  In the coeval 
model every star cluster starts immediately and at the same time to
loose mass and in the delayed model the star clusters start to loose
their mass randomly during the first crossing time of the super-cluster.
While in the coeval cases the star formation rates (SFR) range from
$0.08$ to $0.8$~M$_{\odot}/$yr (all stars form within a crossing time
of a individual cluster) the delayed models imply SFRs from $0.03$ to
$0.3$~M$_{\odot}/$yr (the stars form within the crossing time of the
star cluster complex).  Using the crossing time of the cluster complex
as the maximum time delay between the start of the star formation of
two individual clusters is in agreement with the analysis of
\citet{efr98} and their time difference--separation--relation (see
their fig.~8).

The star cluster complex orbits circularly at a distance of $10$~kpc
around an analytical galactic potential with a flat rotation curve of
$220$~kms$^{-1}$. 

\section{Results}
\label{sec:res}

First we checked if our approach gives the same results on isolated
single clusters as published before by evaluating the remaining bound
mass of an isolated cluster after $1$~Gyr.  The bound mass is computed
by calculating the total energy of each particle at each time step.
If the kinetic energy of the particle with respect to the star cluster
centre exceeds the potential energy derived from the grid-based
potentials of {\sc Superbox} the particle is regarded as being
unbound. 

\begin{figure}[t!]
  \centering
  \plotone{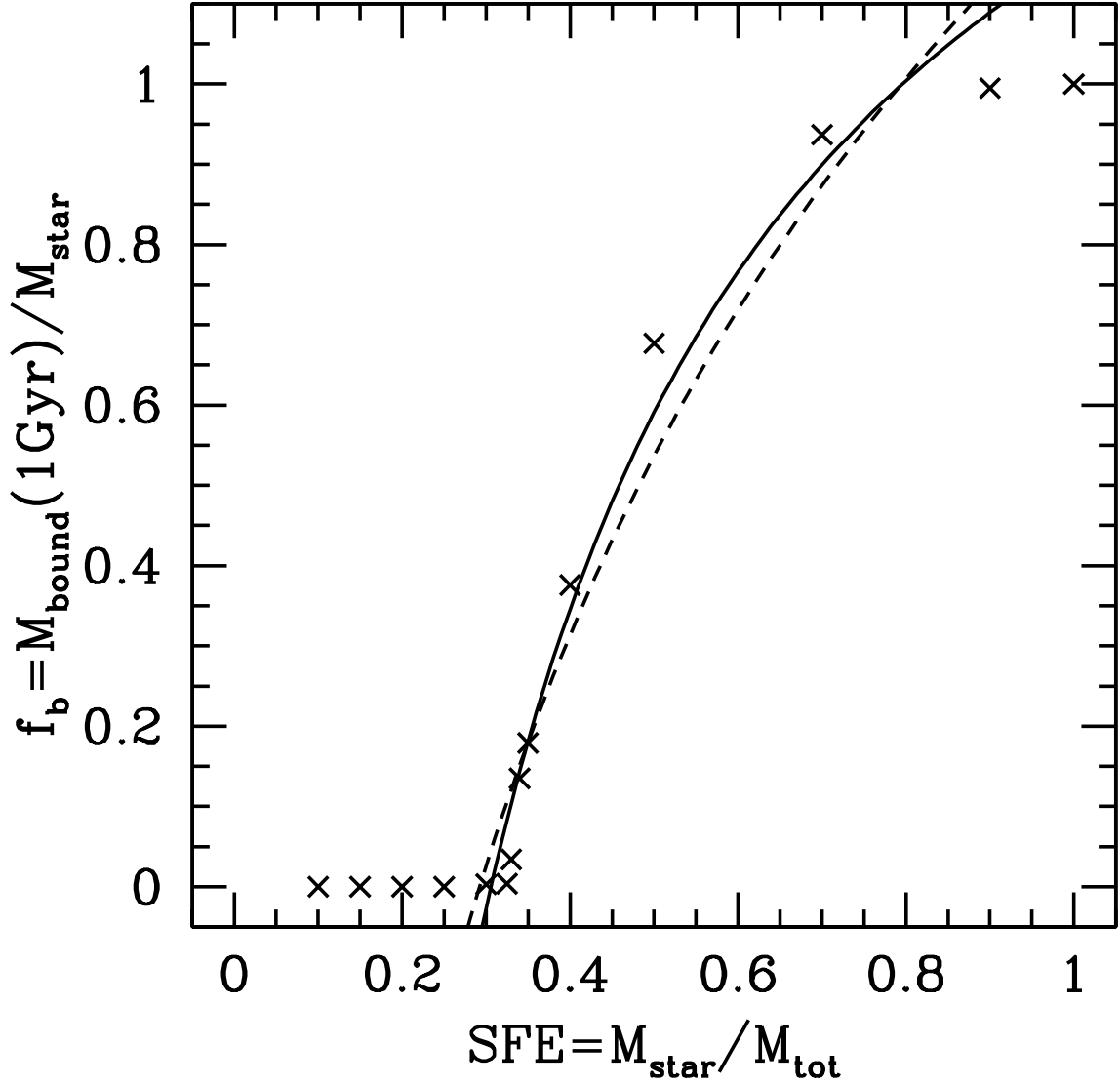}
  \caption{Results of the isolated cluster simulations.  Plotted is
    the bound mass at $t=1$~Gyr divided by the initial mass in stars
    against the star formation efficiency (SFE) which is the initial
    mass in stars divided by the total mass, i.e.\ the mass in stars
    and gas.  Solid line is a power law fit: $f_{b} = -{\rm
      SFE}^{-0.65} + 2.16$; dashed line is a logarithmic fit: $f_{b} =
    \ln({\rm SFE}) + 1.23$.  Only the rising part ($0 < f_{b} < 1.0 $)
    is fitted.}   
  \label{fig:single}
\end{figure}

The results, shown in Tab.~\ref{tab:single} and Fig.~\ref{fig:single},
agree with the numbers in the literature \citep{goo97,boi03a,boi03b}.
If the SFE is higher than $33$~per cent a small bound core survives.
We get a star cluster which retains more than $50$~ per cent of its
stellar mass if the SFE is about $40$--$50$~per cent.  For each SFE
value we performed three simulations with different random
realisations.  The results of the individual realisations all agreed
to better than one per cent.   
 
\begin{figure}[t!]
  \centering
  \plottwo{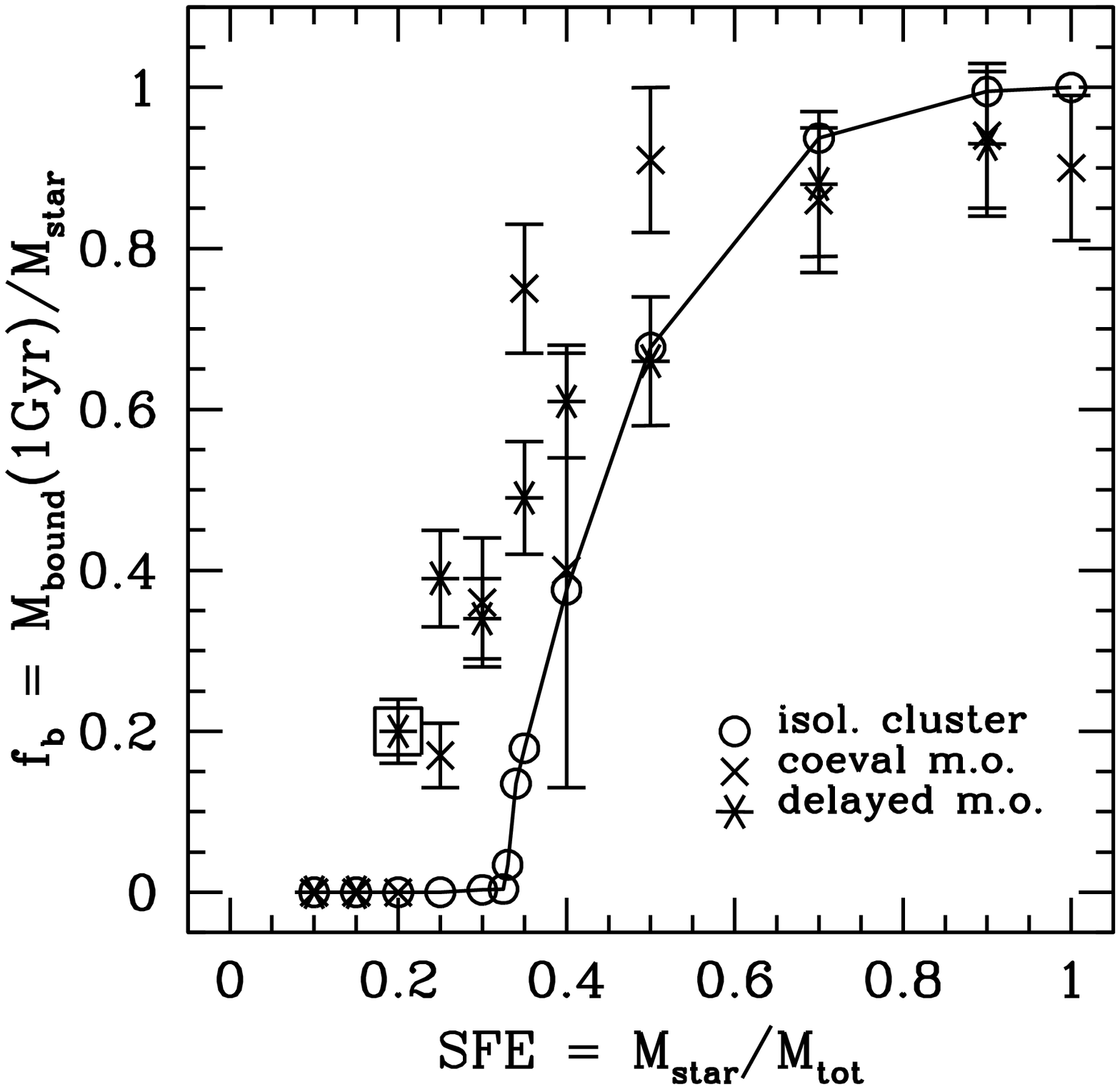}{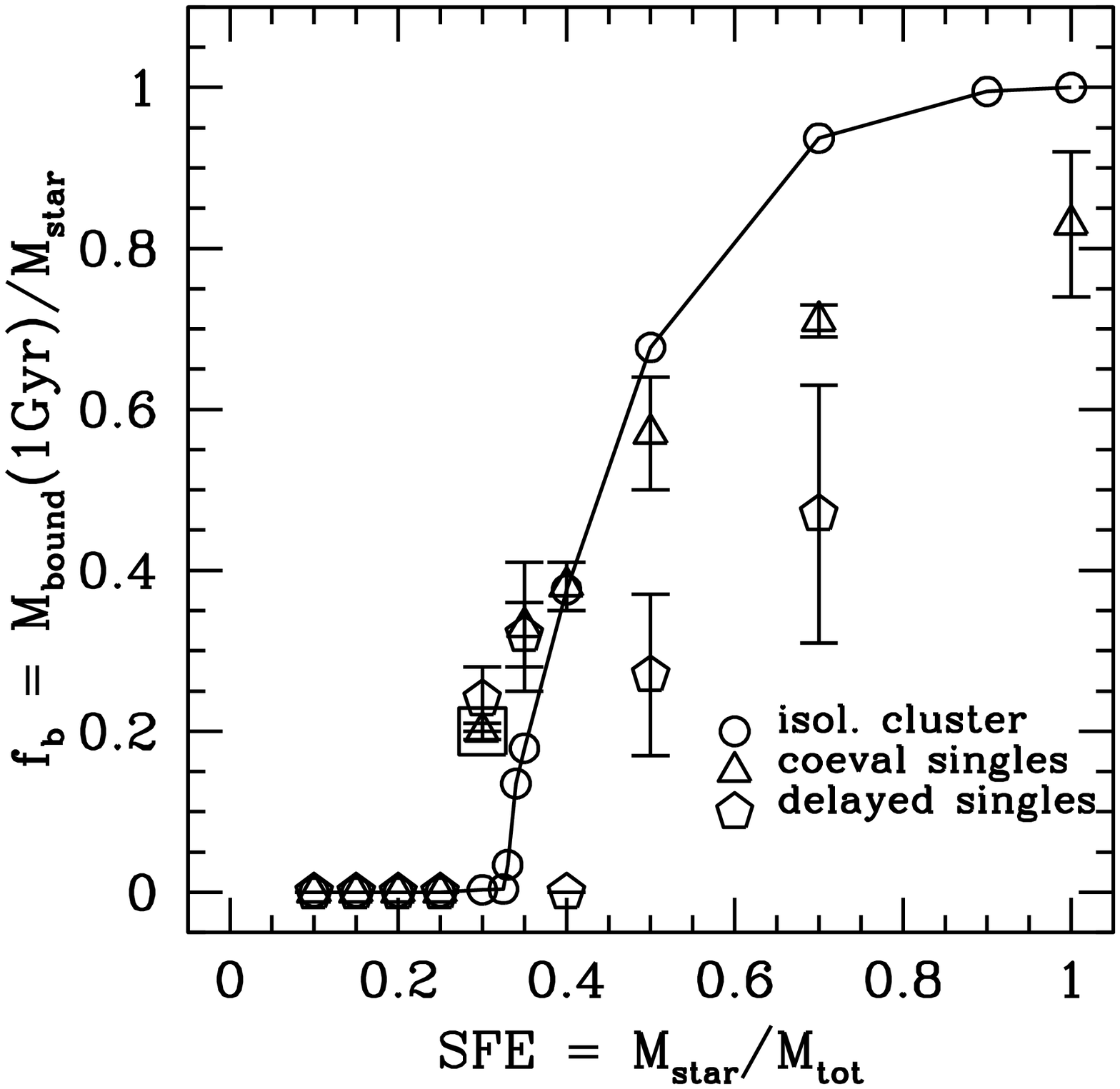}
  \caption{Results of star cluster complex simulations.  Large open
    circles show the results of the isolated clusters
    (Fig.~\ref{fig:single}).  Crosses are the merger objects of the
    simulations with coeval gas-expulsion, six-pointed stars denote
    the merger objects in the simulations with randomly delayed gas
    expulsion. Small open triangles and small open pentagons  
    are surviving, escaped star clusters in the coeval and the delayed
    case respectively. Data points within a large open square denote
    the models discussed in detail in this paper.} 
  \label{fig:multi}
\end{figure}

As a next step we performed simulations with $20$~star clusters as
described above.  In Tab.~\ref{tab:merg} we give the number of star
clusters which have merged (M), the number of clusters which
completely dissolve (D), and the number of star clusters which survive 
as single entities after leaving the potential of the star cluster
complex (S).  For the bound mass fraction of the merger object
($f_{\rm b}^{M}$) we count the merged clusters as well as the
completely dissolved ones.  The mass of the escaped and surviving star
clusters (i.e.\ that have neither dissolved nor merged) is not taken
into account.  If one or more star clusters survive we calculate their 
own bound mass fraction ($f_{\rm b}^{S}$).  The resulting bound mass
fractions are shown in Fig.~\ref{fig:multi}.  Again the values are
taken after $1$~Gyr of evolution.

As a first result of our investigation we find that, in contrast to
the isolated cluster case, the resulting bound mass fractions have a
wide spread.  Even though we just perform one simulation per parameter
set we sometimes find more than one merger object or more than one
surviving star cluster.  In those cases a one sigma deviation can be a
high as $10$~per cent in $f_{\rm b}$.  This can be explained by the
additional but random destructive tidal forces between the star
clusters.  Almost all clusters which do not end up inside the merger
object have a smaller bound mass than in the isolated case.  There are
even star clusters dissolving completely  when the SFE is $70$~per
cent.  On the other hand there are rare cases where single clusters
escape and survive even at low star formation efficiencies.  Two
single clusters escape and survive the coeval simulation with a SFE of
$30$~per cent.  We analyse one of these surviving star cluster in a
sub-section below (Sect.~\ref{sec:m03}).  

The merger objects in our simulations survive with very low SFE.
In the delayed gas expulsion simulations we find a surviving merger
object at a SFE of only $20$~per cent.  We followed the evolution of
the merger object for $10$~Gyr to see if it survives and which
properties it has (Sect.~\ref{sec:d02}).  

Generally speaking the building up of a merger object with its deeper
potential well favours the survival of a bound object that retains
more of its stars than an isolated single cluster would, as long as
the SFE is below $60$~per cent (Fig~\ref{fig:multi}).  If the SFE is
higher destructive processes during the merging process lead to a
mass-loss, i.e.\ stars that are expelled as a result of the kinetic
energy surplus produced during the merging of the clusters (the stars
are then found in the tidal tails), leaving the remaining object with a 
smaller bound mass fraction than the star clusters would have had if
they would have formed in isolation.   

However, as a major result we find that cluster formation in complexes
allows star clusters to survive even if the SFE is as low as $20$~per
cent.  In a dense star cluster complex the crossing time of the star
clusters through the super-cluster, and therefore the merging
time-scale, is short enough that some star clusters have already
merged before they expel their gas.  The much deeper potential wells
of these merger objects are able to retain the stars more effectively
than isolated clusters would. 

\begin{figure}[h!]
  \centering
  \plotone{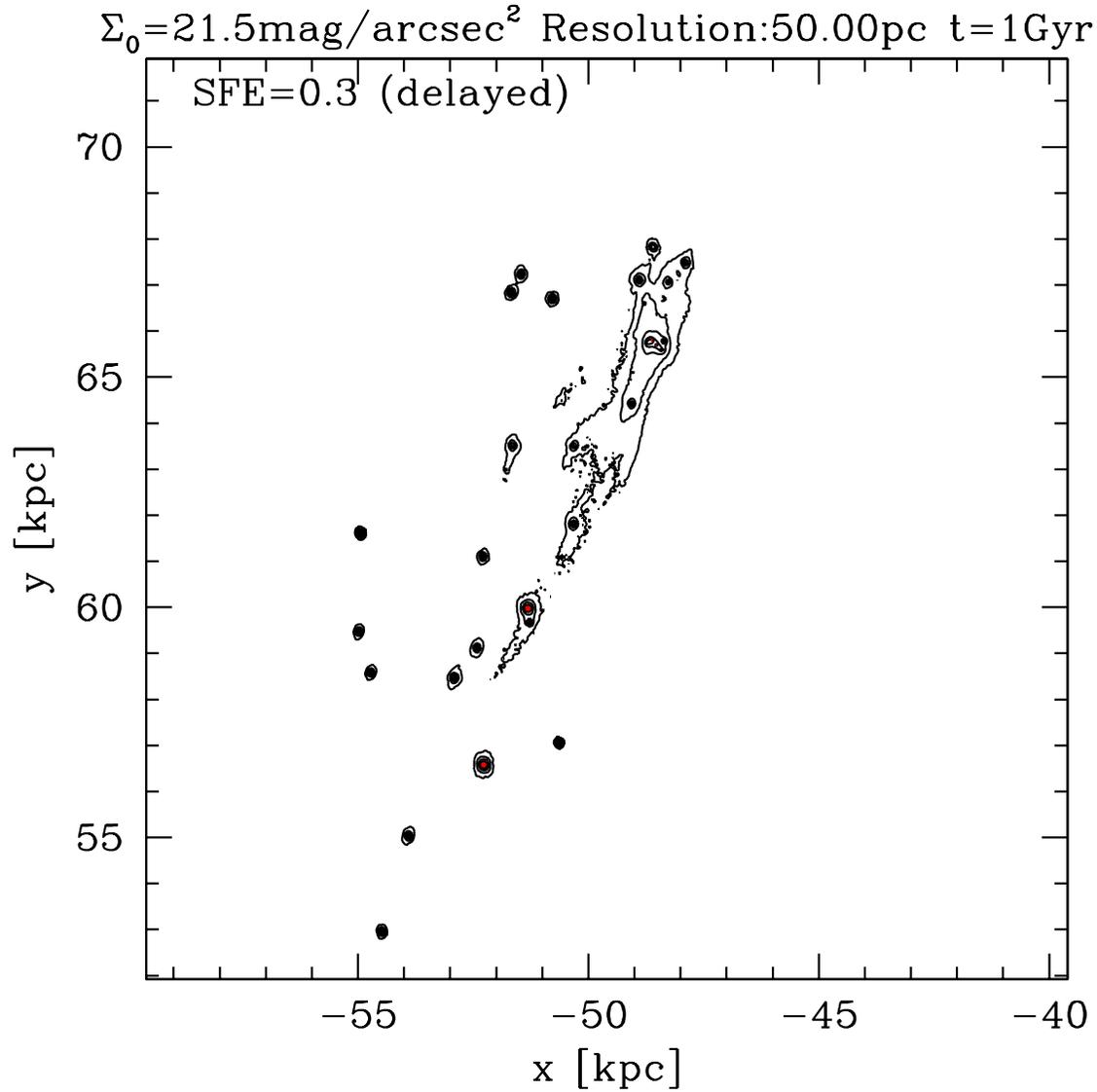}
  \caption{Comparison simulation with a very extended star cluster
    complex (effective radius $= 250$~pc) in the outer halo.  The SFE
    is $30$~per cent and the gas-expulsion is delayed.  Shown in the
    figure is the surface brightness contours of the dispersed star
    cluster complex at $t=1$~Gyr (adopting a mass-to-light ratio of
    $1$).  The surviving single star clusters are clearly visible.} 
  \label{fig:w04}
\end{figure}

To show the influence of the richness of the star cluster complex we
performed one comparison simulation, where we placed an extended star
cluster complex (Plummer-radius of $250$~pc and a cut-off radius of
$1.25$~kpc) on an eccentric orbit far out (apogal: $100$~kpc; perigal:
$80$~kpc).  The complex contains $32$ star clusters and has a total
mass of $8.8 \cdot 10^{6}$~M$_{\odot}$ or $2.64 \cdot
10^{6}$~~M$_{\odot}$ after the delayed gas-expulsion (SFE$=0.3$).
This also implies a low star formation rate of only
$0.02$~M$\odot/$yr.  The crossing-time of this complex is long 
($T_{\rm cr} =124$~Myr) compared to the gas-expulsion time, which
happens on the time-scales of the crossing-time of the single
clusters.  In this simulation only two star clusters merge and two get
dissolved.  All the other star clusters survive but get dispersed,
because the potential of the complex is not deep enough to retain the
star clusters after gas-expulsion.  Also the merging time-scale which
is of the order of a few crossing-times of the complex is too long to
prevent the star clusters from leaving the complex.  Still, as said
above, almost all clusters survive and retain about $29$~per cent of
their initial mass in stars after one Gyr.  A magnitude-spaced
contour-plot of this simulation is shown in Fig~\ref{fig:w04}.

\subsection{Merger Object at a SFE of $20$~per cent}
\label{sec:d02}

In the randomly delayed gas expulsion experiment the merging process
takes place faster than the dissolution of the star clusters.  This
means there is already a massive merger object with a deeper potential
well when the major part of the gas expulsion occurs.  This enables
the object to retain a higher fraction of the stars bound to each
other. 

\begin{figure}[t!]
  \centering
  \plottwo{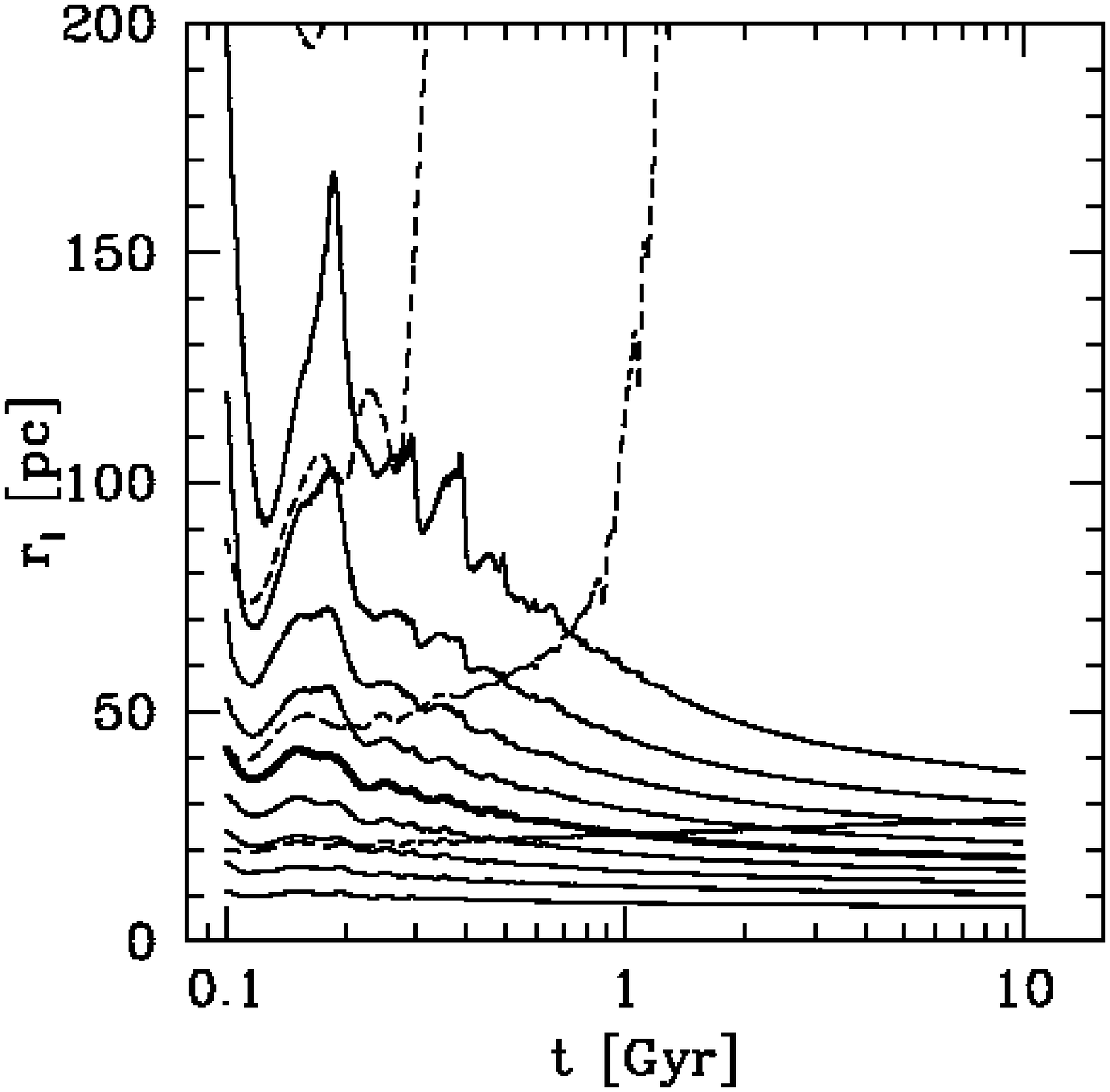}{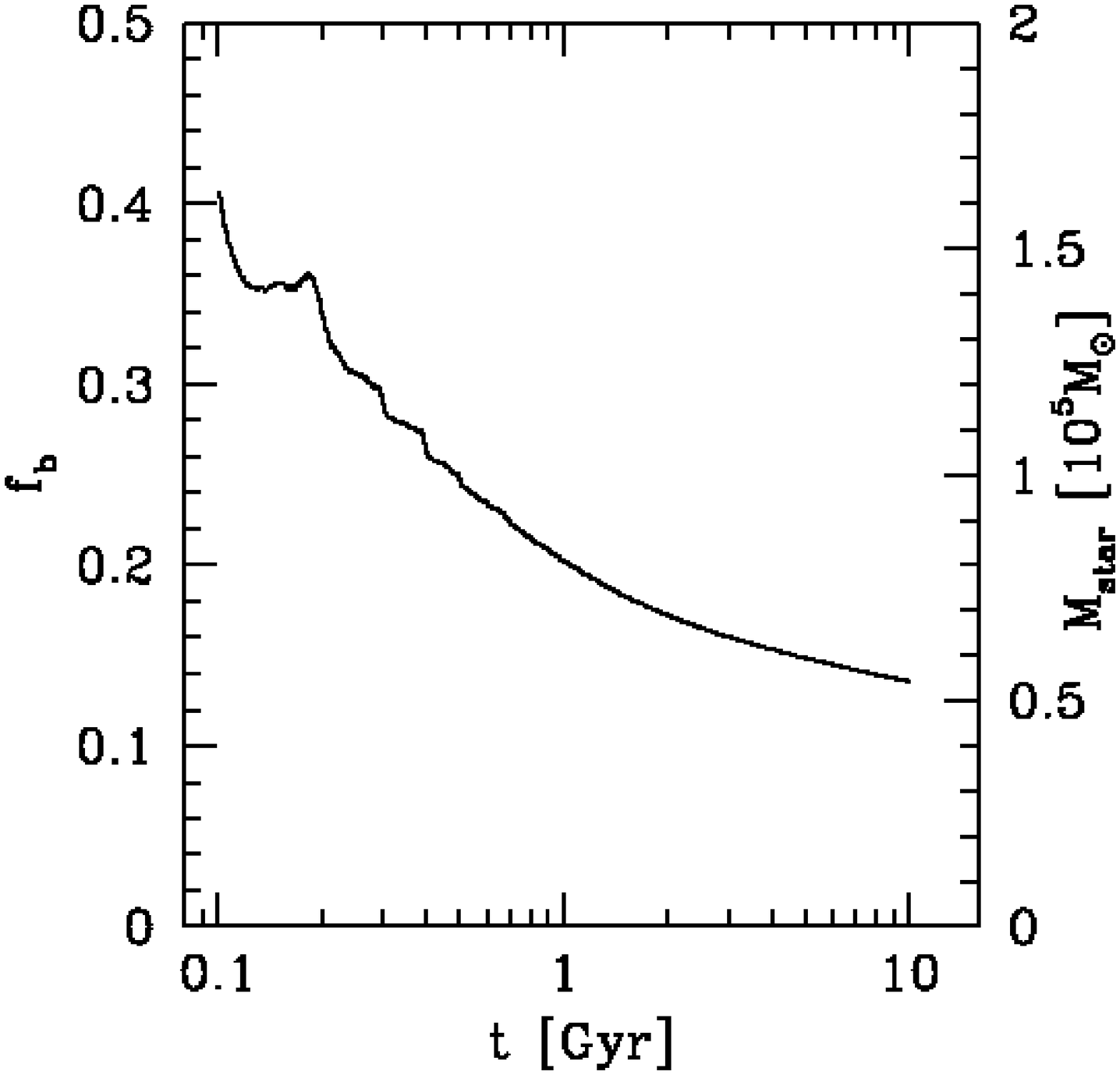}
  \caption{Merger object at a SFE of $20$~per cent (delayed).   Left:
    Lagrangian radii ($10$\%, $20$\%, ..., $90$\%) of all particles
    (dotted) and the bound particles only (solid).  Thick line is the
    half-mass radius.  Right: Fraction of the particles which form a
    bound object.  Right vertical axis shows the mass.  We use an
    unusual logarithmic time-axis in both panels to enhance the
    evolution during the first few hundred Myr.} 
  \label{fig:d02-1}
\end{figure}

We followed the evolution of this object for $10$~Gyr.  After a rapid
mass loss during the first few hundred Myr (Fig.~\ref{fig:d02-1},
right panel) the merger object survives and hosts finally (after
$10$~Gyr) about $14$~per cent of the initially formed stars.  This
amounts in our case to a mass of $5.42 \cdot 10^{4}$~M$_{\odot}$, or,
adopting a typical mass-to-light ratio for globular clusters of $3$,
to an absolute V-band magnitude of $-5.77$.  After a first expansion
phase during the first few dozens of Myr the expansion is halted by
the self-gravity of the object and the particles recollapse again.  It
takes a few hundred Myr until the object settles down but afterwards
it looses mass and shrinks further only slowly (Fig.~\ref{fig:d02-1}).

\begin{figure}[t!]
  \centering
  \plottwo{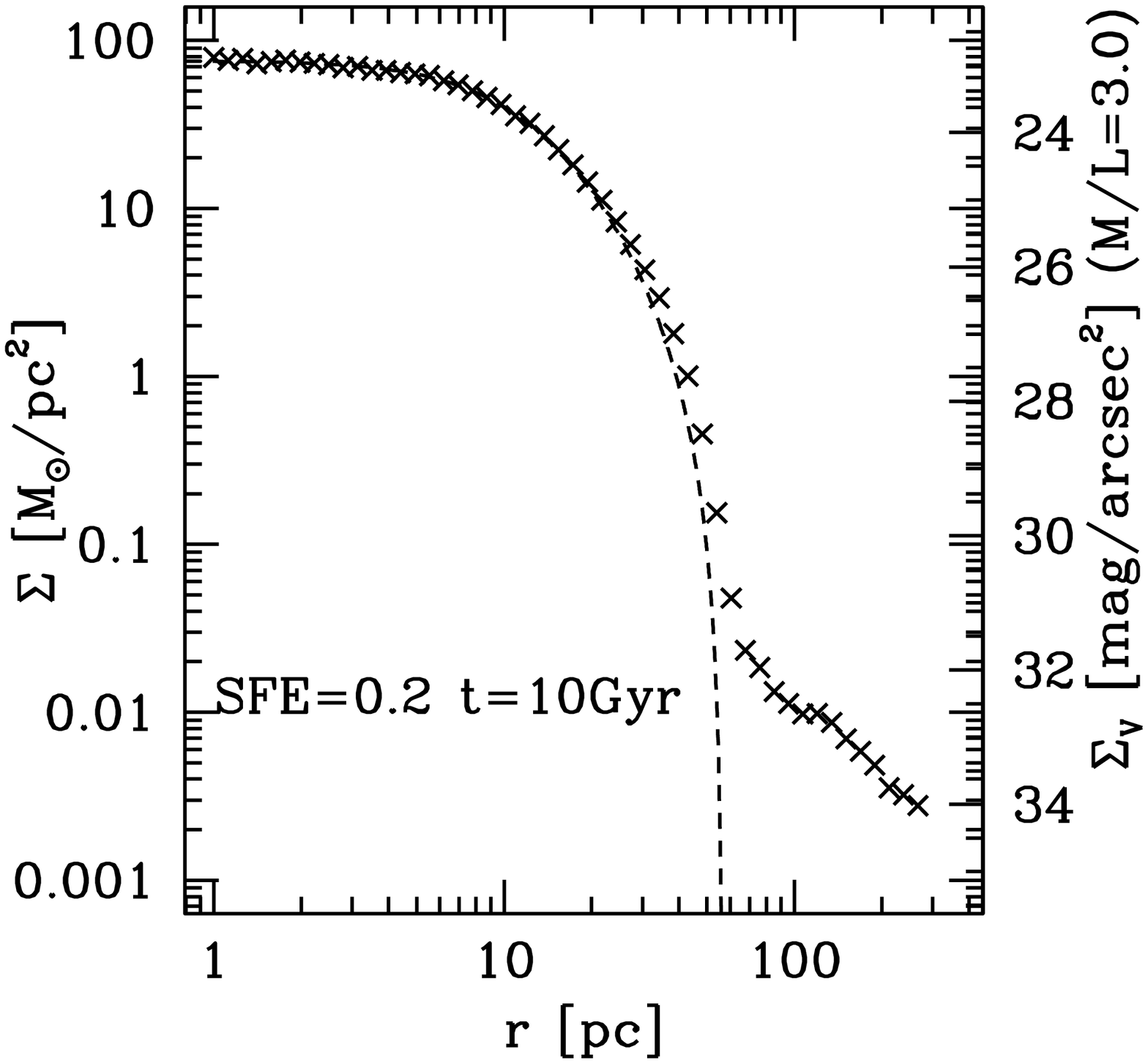}{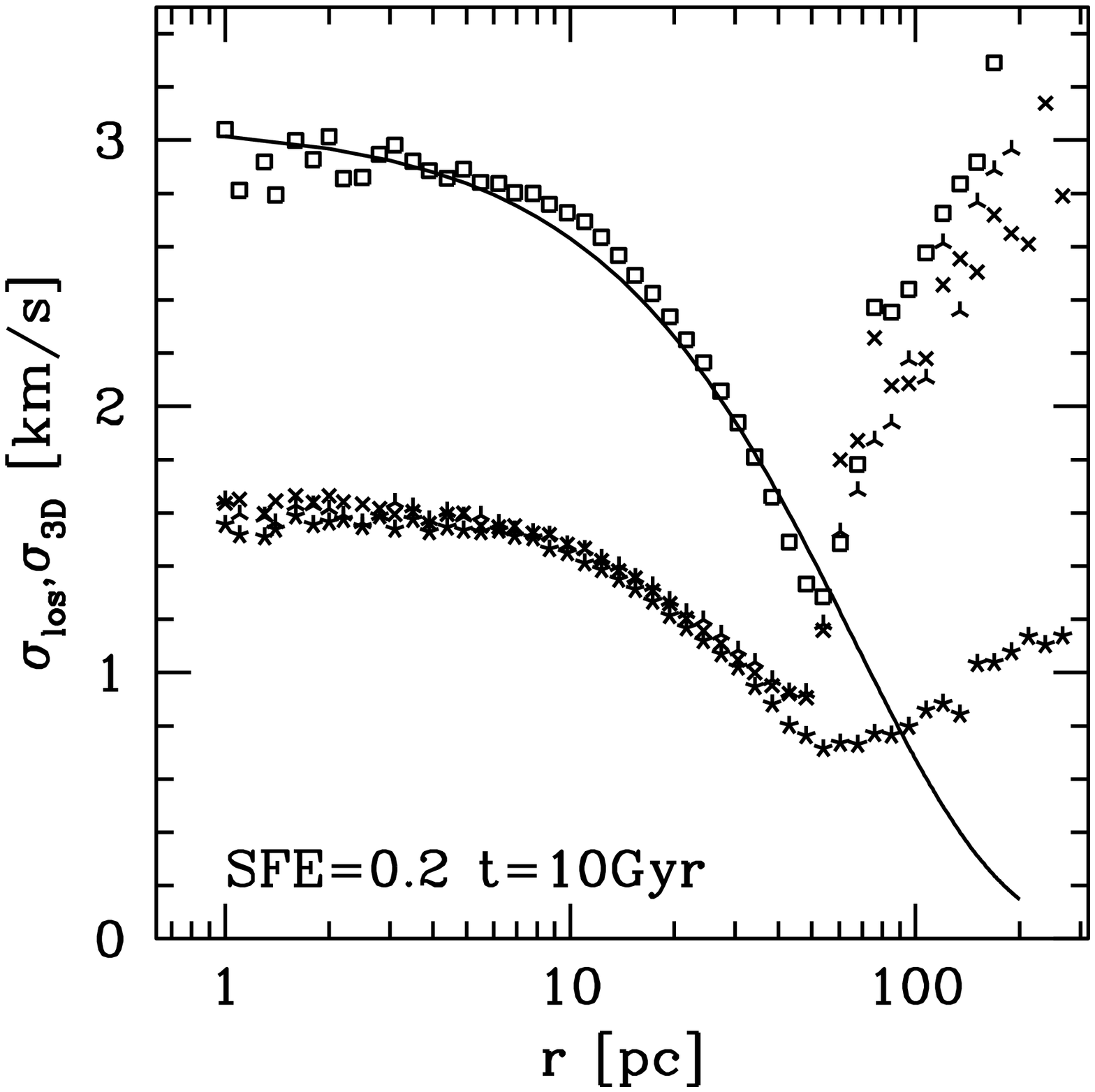}
  \caption{Merger object at a SFE of $20$~per cent after $10$~Gyr of 
    evolution.  Left: Surface density profile.  Crosses are the data 
    points and the solid line is a King profile fit as described in
    the main text.  Right: Velocity dispersions: Tripods, crosses and
    5-pointed stars are the line-of-sight velocity dispersions
    measured in concentric rings around the object in the directions
    of the $x$-, $y$- and $z$-axis respectively.  Open squares are the
    3D-velocity dispersion measured in spherical shells centred on the
    object.  Solid line is an exponential fitting curve as described
    in the main text.} 
  \label{fig:d02-2}
\end{figure}

By $10$~Gyr the object does not look like an 'ordinary' globular
cluster but rather like one of the faint fuzzy star clusters
\citep{lar00}.  It has no dense core and a large effective radius of
about $13$~pc.  The half mass radius of the object is $18.2$~pc
(Fig.~\ref{fig:d02-1} left panel).  We fit a King profile to the
surface density profile with a core radius of $12.8 \pm 0.2$~pc and a
central surface density of $124.8 \pm 0.8$~M$_{\odot}$pc$^{-2}$.  This
translates to $22.4$~mag.arcsec$^{-2}$ (again in the $V$-band) taking a
mass-to-light ratio of $3$.  The tidal radius is $56.5$~pc
(Fig.~\ref{fig:d02-2} left panel).  The velocity dispersion profiles
can be fitted with an exponential profile with a central value of
$1.6$~kms$^{-1}$ and an effective radius of $80$~pc for the
line-of-sight velocity dispersions, and $3.06 \pm 0.02$~kms$^{-1}$ and
$66.1 \pm 2.2$~pc for the three dimensional velocity dispersion.  We
also find a steep rise in the dispersions among unbound particles
outside the tidal radius.  These extra tidal stars are mainly located
in tidal arms leading and trailing the object.

\subsection{Surviving Star Cluster at SFE of $30$~per cent}
\label{sec:m03}

\begin{figure}[t!]
  \centering
  \plottwo{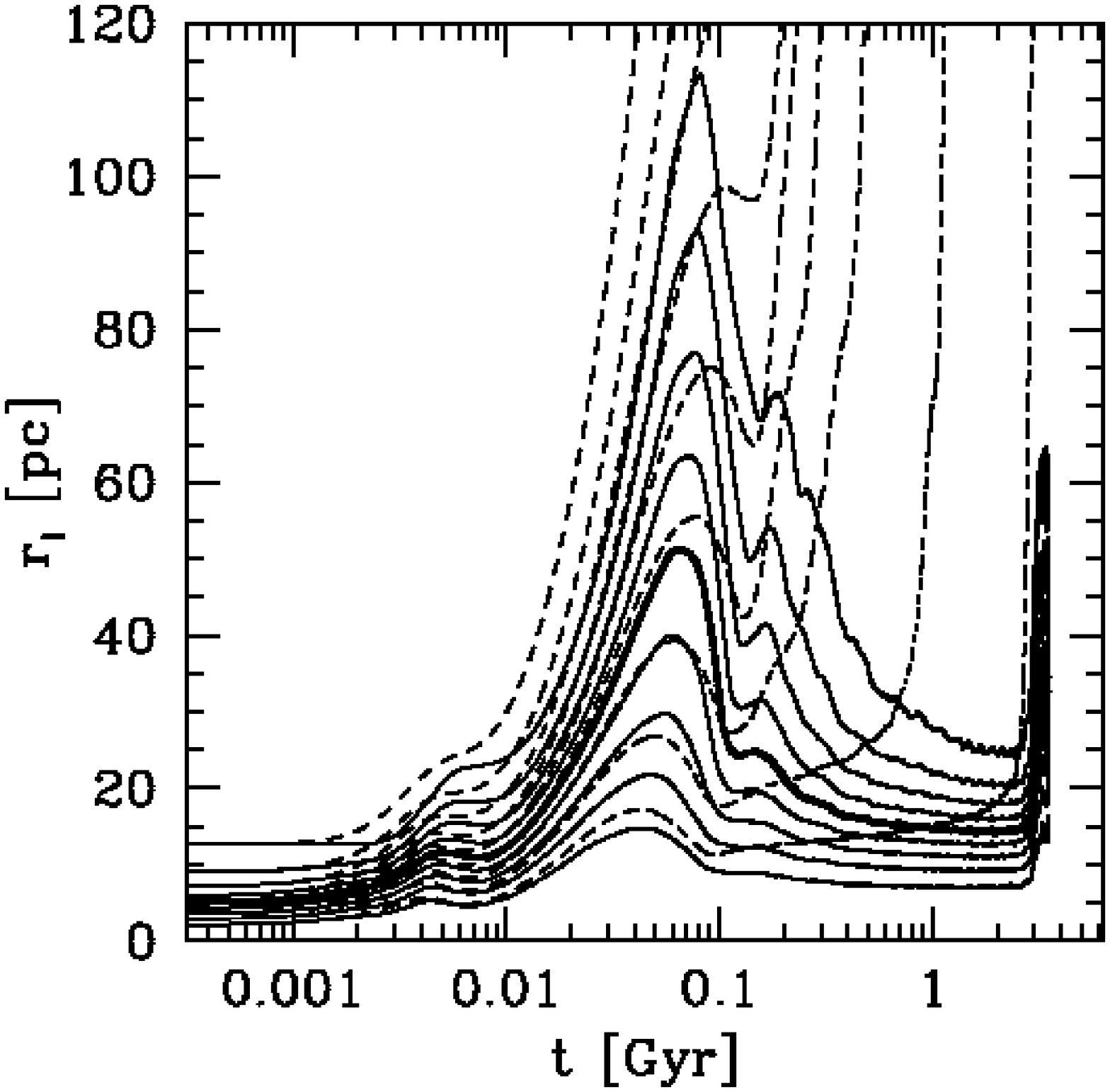}{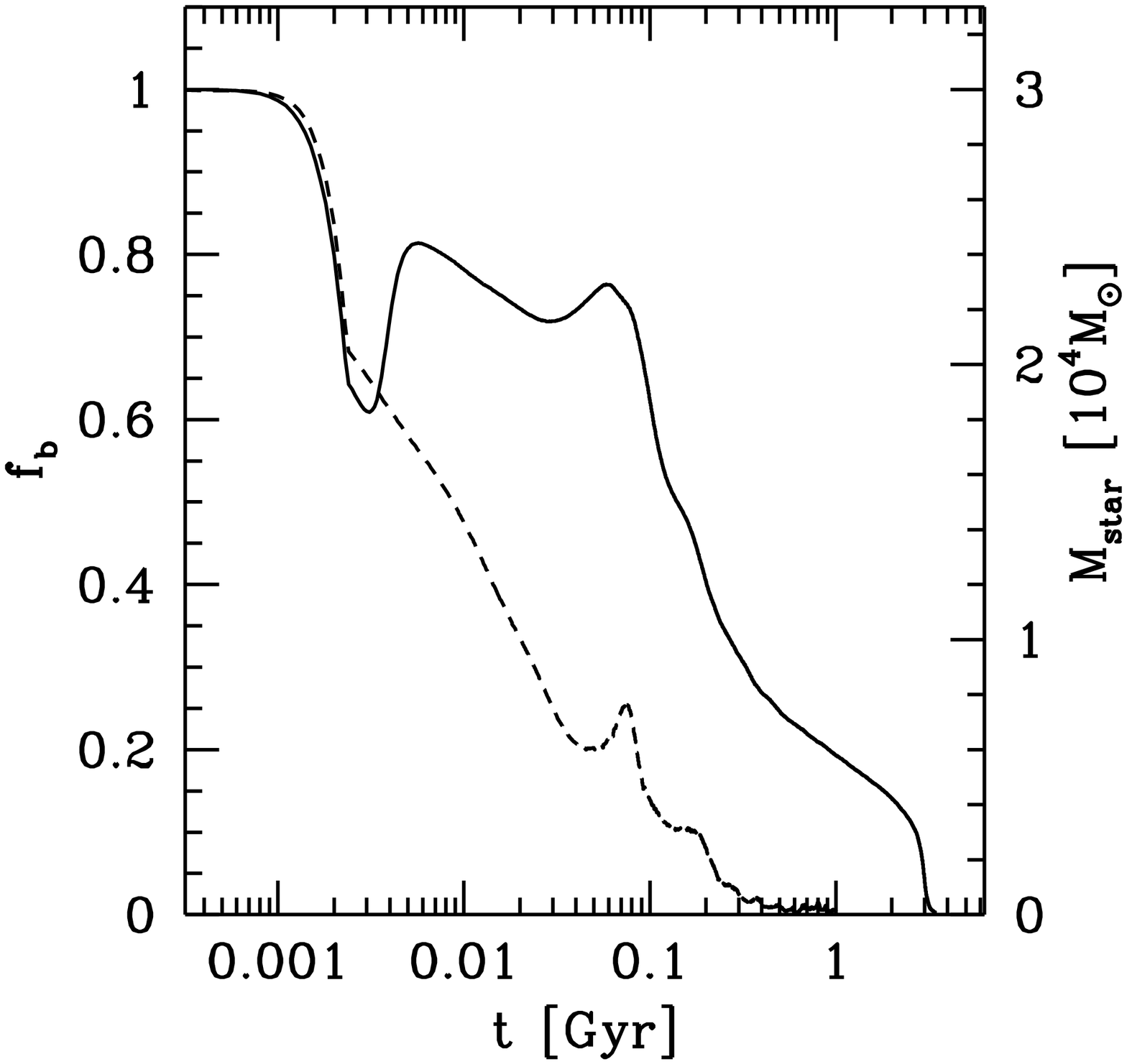}
  \caption{Surviving star cluster at a SFE of $30$~per cent.  Left:
    Lagrangian radii ($10$\%, $20$\%, ..., $90$\%) of all particles
    (dashed lines) and the bound particles only (solid lines).  Thick
    line denotes the half-mass radius.  In this figure the turning
    point after which the remaining bound particles contract again is
    clearly visible.  Right: Bound mass fraction of the surviving star
    cluster (solid line) and for comparison the bound mass of an
    isolated star cluster with the same SFE (dashed line).  Clearly
    visible is the rise in the bound mass after 3~Myr when the star
    cluster gets kicked out of the cluster complex.  Right vertical
    axis denotes the mass.  To highlight the crucial evolution during
    the first few Myr we use time-axes that are logarithmically
    spaced in both panels.}  
  \label{fig:m03-1}
\end{figure}

In the coeval simulation with a SFE of $30$~per cent we find two
single clusters which do not merge and do not dissolve.  They get
kicked out of the potential of the star cluster complex and retain 
$20$~per cent of their stars bound at $t=1$~Gyr.  In the left
panel of Fig.~\ref{fig:m03-1} the bound mass fraction of one of them
is shown. Clearly visible is a steep increase of the bound mass
fraction at about $3$~Myr.  At this time the star cluster has a close
encounter with another star cluster and gets shot out of the cluster
complex.  This event shocked the particle distribution and prevented
otherwise unbound particles from gaining too much energy and becoming
unbound.  In the right panel of Fig.~\ref{fig:m03-1} we see that this
event is almost invisible in the Lagrangian radii of the particles.
They still keep on expanding.  The turning point when the remaining
particles contract again (due to their self-gravity) to form the
surviving object is much later (at about $50$~Myr).  After that the
object evolves and looses mass only slowly thereby shrinking in size
until final dissolution. 

\begin{figure}[t!]
  \centering
  \plottwo{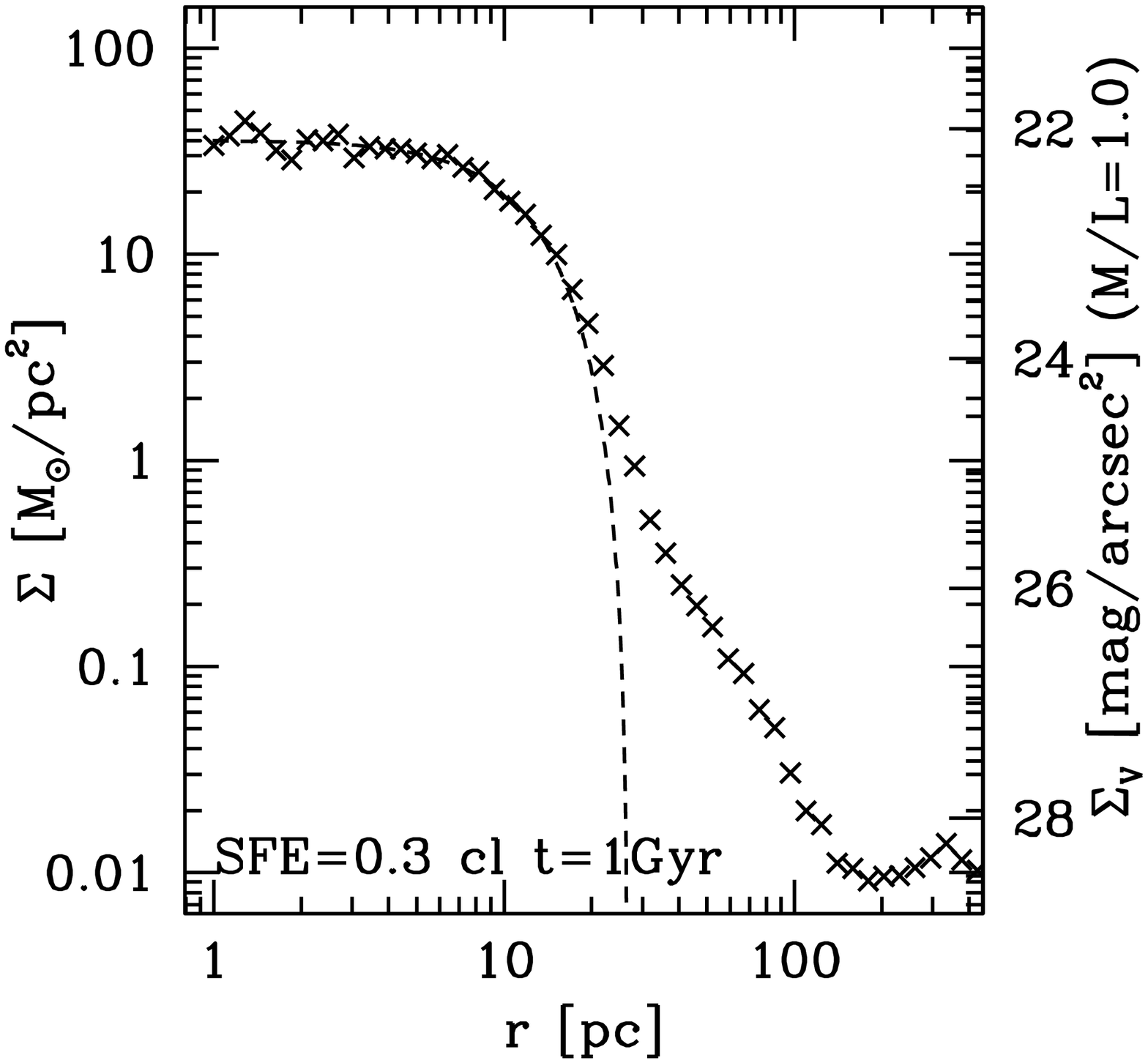}{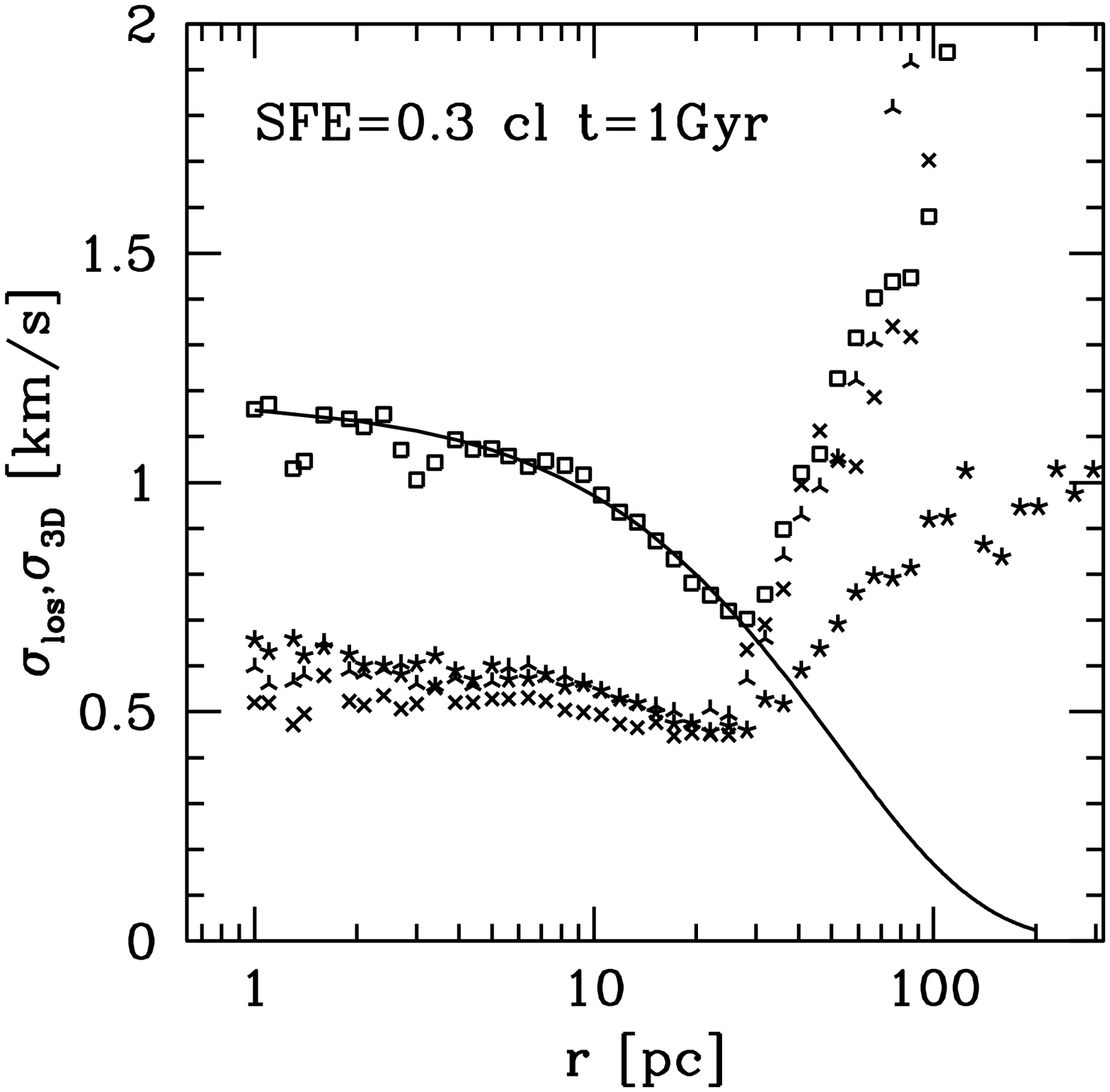}
  \caption{Surviving star cluster at a SFE of $30$~per cent.  Left:
    Surface density profile of a surviving star cluster 
    with a SFE of 30~\% after 1~Gyr. Crosses are the data points and
    dotted line is a King profile fit with a core radius of
    20.6~pc and a tidal radius of 26.8~pc.  Outside the tidal radius
    the shape is a power-law.  Right: Velocity dispersions; Three
    pointed stars, crosses and five pointed stars are the
    line-of-sight velocity dispersions measured in concentric rings
    around the centre of the object along the $x$, $y$- and $z$-axis.
    Open squares are the 3D-velocity dispersion measured in concentric
    shells.}
  \label{fig:m03-2}
\end{figure}

By one Gyr the star cluster has a bound mass fraction of $19$~per
cent which amounts to $5.8 \cdot 10^{3}$~M$_{\odot}$.  The bound
particles form a core surrounded by a halo of unbound particles.  The
core has a tidal radius of about $27$~pc.  The shape of the surface
brightness profile is best fitted with a King profile in the centre
with a core radius of  $20.6$~pc and a power-law in the outer part and
extends further than the tidal radius as shown in the left panel of
Fig.~\ref{fig:m03-2} .  The line-of-sight velocity dispersion has an
exponential shape out to the tidal radius and rises beyond this point
due to the surrounding unbound particles.

To follow the future evolution of less massive objects like this
cluster one would definitely need a direct N-body code to account for
internal evolution caused by two-body relaxation effects.
Nevertheless as shown in Fig~\ref{fig:m03-1} (right panel) with our
code which suppresses two-body relaxation completely the dissolution
time is about $3$~Gyr.  Comparing our result with simulations by
\citet{bau03} (their fig.~3) our result is very similar to that
obtained from their more appropriate simulation (dissolution time
$3.5$~Gyr).  

\section{Discussion and Conclusion}
\label{sec:disc}

In the isolated case our models are able to reproduce the standard
value for the SFE to retain a bound object after gas-expulsion.  The
critical SFE beyond which a few per cent of bound stars are retained
is $33$~per cent (for gas-expulsion within a crossing-time).  The SFE
which leads to an object which one can call a star cluster is about
$50$~per cent.  These results are in excellent agreement with previous
theoretical works \citep{goo97, gey01} using collision-less methods
and give us confidence in the correctness of our approach of treating
gas expulsion.  The use of collisional N-body codes reduces these
values further and experiments \citep{kro01} show a remaining bound
object of size and richness of the Pleiades at a SFE of $30$~per cent
for an initially embedded Orion-Nebula like cluster consisting
of $10^{4}$ stars and brown dwarves.

This work has shown that, for star clusters forming in star cluster
complexes, there is no clear correlation between SFE and final mass of
the object.  This is due to the random realisations of the star
clusters inside the cluster complex leading to different individual
encounter and merger histories.  As a general trend we can state that
this environment is able to assemble a bound object at a much lower
SFE than the isolated case would predict.  We find a surviving bound
object at a star formation efficiency as low as $20$~per cent.  For
low SFEs the resulting bound mass of the remaining object is on
average higher than the isolated case would predict.  But if the SFE
is very high the destructive forces of the encounters and merging
become dominant and the resulting merger objects retain a
significantly lower bound mass than in the isolated case. 

One reason for the ability of cluster complexes to reduce the
necessary SFE for a bound object is the deeper potential
well of the cluster complex which helps to retain some of the escaping
stars.  But another important reason is the merging of the star
clusters.  As already shown in previous studies \citep{fel02} the star
clusters in dense cluster complexes merge on very short time-scales,
namely a few crossing-times of the cluster complex, with a merger
object visible already in the first one or two crossing times.  If the
complex is dense enough, i.e.\ its crossing-time is short and of the
order of the gas-expulsion time-scale, some clusters merge and form a
larger, more massive object before stars react to the gas-loss and try
to escape.  Within the more massive merger object they need a much
higher escape velocity and more stars stay gravitationally bound and
recollapse again to form a bound object.  While such objects can be
dense and massive enough to survive for a Hubble time \citep{bau98},
they do not resemble normal globular clusters because they have rather
large effective radii.  They can only survive as noticeable objects in
a weak tidal field and thus in the outskirts of galaxies, having been
born for example in tidal tails of interacting gas-rich galaxies.  

While our study is based on dense star cluster complexes as found in
the central region in the Antennae \citep{whi99}, we also performed a
simulation with a SFE of $30$~per cent and delayed gas-expulsion
placing an extended cluster complex in the outer halo region.  Even
though, due to the long crossing-time scales of this complex the star
clusters rather disperse than merge, the environment (i.e.\ the deeper
potential of the star cluster complex) is able to help these dispersed
star clusters to survive.  They still resemble bound objects retaining
almost a third of their initial mass in stars even after one Gyr. 

We described the further evolution of one of our merger objects (in
the extreme case) with a SFE of $20$~per cent in detail.  The
properties of this object (low mass, large effective radius) are
similar to those found for the faint fuzzy star clusters in NGC~1023
\citep{lar00}.  Our objects also resemble the faint GCs in the outer
MW halo \citep{cot02, har97}.  The relaxation time of such an object
amounts to $\approx 8.5$~Gyr.  

In one of our simulations we found an escaping and surviving single
star cluster at a SFE which shouldn't allow its survival.  We argued
that a close encounter with another star cluster which lead to an
expulsion instead of a merger was able to 'shock' the particle
distribution such that the expanding and dissolution process was
compensated leaving a low-mass but large, bound star cluster behind.
Even though an object that small would not survive for a Hubble time
due to internal two-body relaxation \citep{bau98} (the relaxation time
measured at $t=1$~Gyr is $2.6$~Gyr, and due to constant mass-loss is
reduced to $800$~Myr at $t=3$~Gyr) one might find young or even
intermediate age objects close to star-burst regions which could 
look like our model.  Another observational counterpart regarding the
physical properties might be the recently discovered \citep{wil04}
unusual 'globular' cluster SDSSJ1049+5103.  It has a half-light radius 
of $23 \pm 10$~pc and a mass in the range of a few hundred up to a
thousand solar masses.  The authors argue that this cluster is either
in the final stage of dissolution, which would make it a perfect
counterpart to our surviving star cluster (seen at its final stage at
$t=3$~Gyr), or a very faint and low-mass dwarf spheroidal galaxy
embedded in a dark matter halo. 

\begin{acknowledgements}
MF thankfully acknowledges financial support through DFG-grant
KR1635/5-1.  
\end{acknowledgements}

\clearpage

\clearpage

\clearpage

\clearpage

\clearpage

\clearpage

\clearpage

\clearpage

\clearpage

\begin{table}[t!]
  \centering
  \caption{Bound mass in per cent after 1 Gyr of evolution. SFE
    denotes the fraction of mass converted into stars; $f_{\rm b}$
    denotes the fraction of stars remaining bound after $1$~Gyr.}
  \label{tab:single}
  \begin{tabular}[t!]{c|llllll} \hline
    SFE & 0.100 & 0.200 & 0.300 & 0.325 & 0.330 & 0.340 \\
    $f_{\rm b}$ & 0.00 & 0.00 & 0.00 & 0.00 & 0.03 & 0.14 \\ \hline
    \hline 
    SFE & 0.350 & 0.400 & 0.500 & 0.700 & 0.900 & \\ 
    $f_{\rm b}$ & 0.18 & 0.38 & 0.68 & 0.94 & 1.00 & \\ \hline 
  \end{tabular}
\end{table}

\clearpage

\begin{table}[t!]
  \centering
  \caption{Results from the star cluster complex simulations. M
    denotes the number of merged clusters, D the number of dissolved
    clusters, and S the number of surviving (unmerged) clusters;
    $f_{\rm b}^{S}$ denotes the bound mass fraction of the surviving
    clusters (if any, otherwise zero) and $f_{\rm b}^{M}$ denotes the
    bound mass fraction of the merger object.  If more than one object
    of either kind is present the numbers denote the mean value.}
  \label{tab:merg}
  \begin{tabular}[t!]{r|rrrrr|rrrrr} \hline
    SFE & \multicolumn{5}{c|}{coeval} & \multicolumn{5}{c}{delayed} \\
    & M & D & S & $f_{\rm b}^{S}$ & $f_{\rm b}^{M}$ & M & D & S &
    $f_{\rm b}^{S}$ & $f_{\rm b}^{M}$ \\ \hline 
    0.10 &  0 & 20 &  0 & 0.00 & 0.00 &  5 & 15 &  0 & 0.00 & 0.00 \\ 
    0.15 &  7 & 13 &  0 & 0.00 & 0.00 & 10 & 10 &  0 & 0.00 & 0.00 \\
    0.20 &  3 & 17 &  0 & 0.00 & 0.00 & 12 &  8 &  0 & 0.00 & 0.20 \\
    0.25 &  7 & 13 &  0 & 0.00 & 0.17 & 14 &  6 &  0 & 0.00 & 0.39 \\
    0.30 & 13 &  5 &  2 & 0.20 & 0.36 & 12 &  7 &  1 & 0.24 & 0.34 \\
    0.35 & 14 &  3 &  3 & 0.33 & 0.75 & 10 &  8 &  2 & 0.32 & 0.49 \\
    0.40 & 14 &  3 &  3 & 0.38 & 0.40 & 13 &  7 &  0 & 0.00 & 0.61 \\
    0.50 & 19 &  0 &  1 & 0.57 & 0.91 & 15 &  3 &  2 & 0.27 & 0.66 \\
    0.70 & 18 &  0 &  2 & 0.71 & 0.86 & 17 &  1 &  2 & 0.47 & 0.88 \\
    0.90 & 20 &  0 &  0 & --   & 0.94 & 20 &  0 &  0 & --   & 0.93 \\
    1.00 & 19 &  0 &  1 & 0.83 & 0.90 & -- & -- & -- & --   & -- \\
    \hline 
  \end{tabular}
\end{table}


\begin{thebibliography}{}

\bibitem[Baumgardt (1998)]{bau98} Baumgardt, H., 1998, \aa, 330, 480

\bibitem[Baumgardt \& Makino (2003)]{bau03} Baumgardt, H., Makino, J.,
  2003, \mnras, 340, 227

\bibitem[Boily \& Kroupa (2003a)]{boi03a} Boily, C.M., \& Kroupa, P.,
  2003, \mnras, 338, 665 

\bibitem[Boily \& Kroupa (2003b)]{boi03b} Boily, C.M., \& Kroupa, P.,
  2003, \mnras, 338, 673  

\bibitem[Burkert \& Bodenheimer (2000)]{bur00} Burkert, A., \&
  Bodenheimer, P., 2000, \apj, 543, 822

\bibitem[Clark \& Bonnell (2004)]{cla04} Clark, P.C., \& Bonnell,
  I.A., 2004, \mnras, 347, L36

\bibitem[C\^{o}t\'{e} et al. (2002)]{cot02} C\^{o}t\'{e}, P.,
  Djorgovsksi, S.G., Meylan, G., Castro, S., \& McCarthy, J.K., 2002,
  \apj, 574, 783

\bibitem[Efremov \& Elmegreen (1998)]{efr98} Efremov, Y.N., \&
  Elmegreen, B.G., 1998, \mnras, 299, 588

\bibitem[Fellhauer et al. (2000)]{fel00} Fellhauer, M., Kroupa, P.,
  Baumgardt, H., Bien, R., Boily, C.M., Spurzem, R., \& Wassmer, N.,
  2000, \na, 5, 305

\bibitem[Fellhauer et al. (2002)]{fel02} Fellhauer, M., Baumgardt,
  H., Kroupa, P., \& Spurzem, R., 2002, Cel.Mech.\& Dyn.Astron., 82, 113 

\bibitem[Fellhauer \& Kroupa (2002a)]{fel02a} Fellhauer, M., \& Kroupa,
  P. 2002, \mnras, 330, 642 
  
\bibitem[Fellhauer \& Kroupa (2002b)]{fel02b} Fellhauer, M., \&
  Kroupa, P. 2002, \aj, 124, 2006 
  
\bibitem[Geyer \& Burkert (2001)]{gey01} Geyer, M.P., Burkert, A.,
  2001, \mnras, 323, 988 

\bibitem[Goodwin (1997)]{goo97} Goodwin, S., 1997, \mnras, 284, 785

\bibitem[Harris (1997)]{har97} Harris, W.E., 1997, VizieR Online Data
  Catalogue, 7202, 0

\bibitem[Johnstone et al. (2000)]{joh00} Johnstone, D., Wilson, C.D.,
  Moriaty-Schieven, G., Joncas, G., Smith, G., Gregersen, E., \& Fich,
  M., 2000, \apj, 545, 327

\bibitem[Klessen \& Burkert (2001)]{kle01} Klessen, R.S., \& Burkert,
  A., 2001, \apj, 549, 386

\bibitem[Kroupa (1998)]{kro98} Kroupa, P., 1998, \mnras, 300, 200

\bibitem[Kroupa et al. (2001)]{kro01} Kroupa, P., Aarseth, S., \&
  Hurley, J., 2001, \mnras, 321, 699

\bibitem[Lada \& Lada (2003)]{lad03} Lada, C.J., \& Lada, E.A., 2003,
  ARA\&A, 41, 57

\bibitem[Larsen \& Brodie (2000)]{lar00} Larsen, S.S., \& Brodie,
  J.P. 2000, \aj, 120, 2938  

\bibitem[Larsen et al. (2002)]{lar02} Larsen, S.S., Efremov, Y.N.,
  Elmegreen, B.G., Alfaro, E.J., Battinelli, P., Hodge, P.W., \&
  Richtler, T. 2002, \apj, 567, 896

\bibitem[Leitherer et al. (1999)]{lei99} Leitherer, C., Schaerer, D.,
  Goldader, J.D., Delgado, R.M.G., Robert, C., Kune, D.F., de Mello,
  D.F., Devost, D., Heckman, T.M. 1999, ApJS, {\bf 123}, 3

\bibitem[MacLow \& Klessen (2004)]{mac04} MacLow, M., \& Klessen, R.,
  2004, RvMP, 76, 125

\bibitem[Tilley \& Pudritz (2004)]{til04} Tilley, D.A., \& Pudritz,
  R.E., 2004, \mnras, 353, 769

\bibitem[Whitmore et al. (1999)]{whi99} Whitmore, B.C., Zhang, Q.,
  Leitherer, C., \& Fall, S.M., 1999,   \aj, 118, 1551

\bibitem[Willman et al. (2004)]{wil04} Willman, B., Blanton, M.R.,
  West, A.A., Dalcanton, J.J., Hogg, D.W., Schneider, D.P., Wherry,
  N., Yanny, B., \& Brinkman, J., 2004, \aj submitted, astro-ph/0410416 

\bibitem[Zhang \& Fall (1999)]{zha99} Zhang, Q., \& Fall, S.M., 1999,
  \apj, 527, 81

\end{thebibliography}
\end{document}